\documentstyle[12pt]{article}
\begin{document}
\thispagestyle{empty}
$\phantom{x}$\vskip 0.618cm\par
{\huge \begin{center}Chiral Decomposition For Non-Abelian Bosons
\end{center}}\par
%\vfill
\begin{center}
$\phantom{X}$\\
{\Large Nelson R.F. Braga and Clovis Wotzasek}\\[3ex]
{\em Instituto de F\'\i sica\\
Universidade Federal do Rio de Janeiro\\
21945, Rio de Janeiro, Brazil\\}
\end{center}\par
%\vfill
 
%\maketitle
\abstract

\noindent 
We study the non-abelian extension for the splitting of a scalar
field into chiral components.  Using this procedure we find a non
ambiguous way of coupling a non abelian chiral scalar field to
gravity.  We start with a (non-chiral) WZW model covariantly coupled
to a background metric and, after the splitting, arrive at two chiral
Wess-Zumino-Witten (WZW) models coupled to gravity.  %\vspace{2cm}
%\noindent PACS:

\newpage %\section{Introduction}
\noindent{\bf 1}   There are indications that a deeper understanding of such
issues as string dynamics and fractional quantum Hall effect phenomenology
can be achieved by treating the chiral
sectors in a more independent way.  However, coupling chiral fields
to external gauge and gravitational fields is problematic.  In a
recent paper\cite{BW}, we have discussed how the coupling of chiral
(abelian) fields to external gravitational backgrounds can be
achieved by diagonalization of the first-order form of a covariant
scalar action.  The theory reduces then to a sum of a left and a
right Floreanini-Jackiw actions\cite{FJ}, circumventing the problems
caused by the lack of manifest Lorentz invariance.  Proceeding along
these lines, one can gauge the scalar action before chiral splitting,
which is trivial, and thus obtain the correctly gauged chiral scalar
action. This procedure was  motivated by a previous article by
Bastianelli and van Nieuwenhuizen\cite{BN} where the coupling of one
abelian chiral boson to gravity is achieved by starting with scalar
field coupled covariantly to gravity and then imposing a chiral
constraint in the first order Lagrangean.

In this Letter we extend the chiral decomposition scheme to nonabelian
scalar fields and use this result to study the coupling
with a gravitational background field.
To introduce the basic ideas and some notation, 
we will briefly present the results of Tseytlyn and West\cite{TW} (TW), where
the scalar field action is written in terms of dual symmetric variables.  This
formulation of the theory provides a natural and adequate framework for the
discussion of chiral dynamics, in the abelian case.  We
will see that the TW dual action can be easily diagonalized when written in
terms of chiral variables, including the case of gravitational coupling. 
Although such description in terms of dual variables does not survive
the non-abelian extension, we will see that it is still possible to
define chiral variables and obtain a diagonalized action.  To this
end we study the Principal Chiral Model (PCM), in its 1st-order form,
and show that it cannot be diagonalized following the above
procedure.  However, the Wess-Zumino topological term produces, under
certain conditions, a canceling piece for the non diagonal part of
PCM, resulting into two Sonnenchein's chiral bosons\cite{S} of
opposite chiralities.  To consider the gravitational coupling, and
show that beginning with the minimal coupled WZW model and then
splitting it in terms of the chiral fields leads to right and left
non abelian chiral bosons coupled to gravity.

\vspace{0.3cm}

%\section{Chiral decomposition for the abelian field}

\noindent{\bf 2}    The separation of a scalar field into its chiral
components has been
introduced by Mandelstan\cite{Mandelstan} in its seminal paper on 2D
bosonization.  The chiral splitting for the
non-abelian side has been studied by Polyakov and Wiegman\cite{PW}.  
However, the abelian limit of Polyakov-Wiegman decomposition does not
coincide with Mandelstan chiral decomposition.  The chiral separation
in \cite{Mandelstan} is based on a first-order theory
while that of \cite{PW} is second-order. Mandelstan chiral
decomposition scheme can be obtained as a constraint over the theory
requiring complete separability of the original action.  To see how
this comes around we write the scalar action in terms of
Tseytlin-West dual variables $\varphi$ and $\tilde\varphi$:

\begin{eqnarray}
\label{1st-order}
{\cal S}(\varphi,\tilde\varphi)&=&{1\over 2}\left({\tilde\varphi^\prime}
{\dot\varphi} +\dot{\tilde\varphi}\varphi^\prime-{\tilde\varphi^\prime}
{\tilde\varphi^\prime}-
{\varphi^\prime}{\varphi^\prime} \right)\nonumber\\
&=&  {1\over 2} \Phi^T\partial_\sigma\left(I \partial_\sigma  -
\sigma_1\partial_\tau\right)\Phi
\end{eqnarray}

\noindent where $\Phi^T=\left(\varphi,\tilde\varphi\right)$,
$\sigma_k$ are Pauli matrices, and $I$ is the identity matrix.  Here
dot and prime have their usual significance as time ($\partial_\tau$)
and space ($\partial_\sigma$) derivatives.  Notice that this action
is invariant under the interchange of
$\varphi\leftrightarrow\tilde\varphi$.  From their field equations,
plus some appropriated boundary conditions\cite{TW}, one can see that
they are dual to each other

\begin{equation}
\label{dualidade}
\partial_a\varphi = \epsilon_{ab}\partial^b\tilde\varphi\;\;\; .
\end{equation}

\noindent  The basic idea in \cite{BW} is to make a change of variables,
say $(\varphi,\tilde\varphi)\rightarrow (A,B)$ such that the first-order,
self-dual action (\ref{1st-order}) decouples as a sum of the individual
actions for each field: ${\cal S}(\varphi,\tilde\varphi)\rightarrow 
{\cal S}(A,B)= S(A)+S(B) $.  This intent is achieved if we define the
Mandelstan chiral variables
as a linear combination of the original dual variables.  The separability
condition then gives the following combination:

\begin{eqnarray}
\label{redefinition}
\varphi&=& A + B\nonumber\\
\tilde\varphi&=& \varepsilon\left(A - B\right)
\end{eqnarray}

\noindent where $\varepsilon=\pm 1$ is the only freedom remaining.  One can
recognize the set $(A,B)$ as the chiral basis, where the role of
$\varepsilon$ is to interchange the chirality of the individual components. 
Indeed, the transformation (\ref{redefinition}) from the dual basis
to the chiral basis above diagonalizes the matricial structure in the
action (\ref{1st-order})

\begin{equation}
\left(\partial_\sigma  I -\partial_\tau\sigma_1\right)\rightarrow 
\partial_\sigma I
-\varepsilon\partial_\tau\sigma_3
\end{equation}

\noindent  so that the final action reads

\begin{eqnarray}
\label{scalarmandelstan}
{\cal S}  &=&\int d^2x\left[ \left(\varepsilon A^{\prime}\dot A-
A^{\prime}A^{\prime}\right)
 +\left(-\varepsilon B^{\prime}\dot B - B^{\prime}B^{\prime}\right)\right]
\nonumber\\
&=& {\cal S}^{(+\varepsilon)}\left(A\right)+
{\cal S}^{(-\varepsilon)}\left(B\right)
\end{eqnarray}

\noindent which represents two chiral bosons in the formulation of
Floreanini and Jackiw\cite{FJ}, respectively a left and a right moving
boson for $\varepsilon=1$, and right and a left going one for
$\varepsilon =-1$:

\begin{eqnarray}
(i)\:\:\:\:\:\:\varepsilon &=& 1\nonumber\\
A(x) &=& \varphi_+(x^+)+ h_+(t)\nonumber\\
B(x) &=& \varphi_-(x^-)+h_-(t)
\end{eqnarray}

\begin{eqnarray}
(ii)\:\:\:\:\:\:\varepsilon &=& -1\nonumber\\
A(x) &=& \varphi_-(x^-)+h_-(t)\nonumber\\
B(x) &=& \varphi_+(x^+)+h_+(t)
\end{eqnarray}

\bigskip

\noindent where $x^{\pm} = (x^0 \pm x^1 )/\sqrt{2}\,\,$. We observe that the
only ambiguity left in the Mandelstan decomposition, the sign
$\varepsilon$, has no physical significance here, but has an
implication in the non-abelian case.  Finally, we mention that since
$A$ and $B$ are (chiral) components of the same multiplet, we must
have that the time dependent functions $h_-(t)+h_+(t)+0$ as a
constraint.  This is so because these symmetries are not present in
the original theory.  This indicates that arbitrary variations in one
chirality are known by the other chirality through this constraint.

This procedure can be generalized to more general settings. 
Take for instance the action for a scalar field under a background gravity

\begin{equation}
\label{scalargrav}
S = {1 \over 2}\int d^2x \sqrt{-\eta} \eta^{\mu\nu} \partial_{\mu} \phi
\partial_{\nu}\phi
\end{equation}

\noindent In terms of TW dual variables the action
(\ref{scalargrav}) reads
 
\begin{equation}
\label{scalarfirstgrav}
{\cal L}={1\over 2}\left[\tilde\varphi^\prime\dot\varphi +
\dot{\tilde\varphi}\varphi^\prime
-{1\over{2\sqrt{-\eta}\eta^{00}}}\left({{\tilde\varphi}^\prime}
{{\tilde\varphi}^\prime} 
+{{\varphi}^\prime}{{\varphi}^\prime}\right)
+ {{2 \eta^{01}}\over{\eta^{00}}}{\tilde\varphi}^\prime\varphi^\prime\right]
\end{equation}

\noindent This action is clearly invariant under the interchange
$\varphi\leftrightarrow\tilde\varphi$.  From their field equations
one can show that they are dual to each other, in the following sense

\begin{equation}
\partial_a\varphi=\sqrt{-\eta}\epsilon_{ab}\eta^{bc}\partial_c\tilde\varphi
\end{equation}

\noindent Introducing the chiral (Mandelstan) components as in
(\ref{redefinition}),we can generalize the result of \cite{TW} and
get two Floreanini-Jackiw chiral bosons 
coupled to gravity:

\begin{equation}
\label{chiralgrav}
S= S_{\varepsilon}(A,{\cal G_\varepsilon}) + S_{-\varepsilon}
(B,{\cal G_{-\varepsilon}})
\end{equation} 

\noindent where the interacting FJ chiral action is

\begin{equation}
\label{chiralgrav2}
S_\varepsilon(X)=\int d^2x \left(\varepsilon \dot{X}X^\prime-
{\cal G_{\varepsilon}} X^\prime X^\prime\right)
\end{equation}

\noindent and the couplings are given by:

\begin{equation}
\label{couplings}
{\cal G_{\varepsilon}}={1 \over \eta^{00}}\left({1 \over
\sqrt{-\eta}}-\varepsilon \eta^{01}\right)
\end{equation}

\noindent  Since this decomposition does not mix the right and left
components, it provides an unambiguous procedure for the coupling of
chiral bosons to gravity.\vspace{0.3cm}\\

%\section{Chiral decomposition for the non abelian field}

\noindent{\bf 3}    Let us next extend the separability condition discussed
above to non-abelian bosons.  The most obvious choice would be to
consider an action given by a bilinear gradient of a matrix-valued
field $g$ taking values on some compact Lie group $G$, which would be
the natural extension of the free scalar abelian field.  This is the
action for the principal chiral model that reads

\begin{equation}
\label{pcm}
{\cal S}_{PCM}(g)= {1\over 2} \int d^2 x tr \left( 
\partial_\mu g \;\partial^\mu g^{-1} \right) 
\end{equation}

\noindent  Here $g:R^{1,1}\rightarrow G$ is a map from the 2 dimensional
Minkowski space-time to $G$. 
This action, however, puts some difficulties.  First of all, by examining its
field equation we learn that unlike (\ref{1st-order}), it does not
represent a free field.  More important for us is the fact that there
is no simple way to represent this action in terms of TW dual
variables, as in the abelian case.  Surprisingly though, we learn
that it is still possible to introduce chiral variables in a simple
fashion, after the reduction to a first-order action.  The reason
being that chiral variables only appear as derivatives, unlike the
case of dual variables.  The Jacobian of the field redefinition does
not involve a time-derivative and can be reabsorbed in the
normalization of the partition function.  Let us write the PCM in its
first-order form as

\begin{eqnarray}
\label{WZW1}
{\cal S}_{PCM}(g,P) & = & { 1\over 2 } \int d^2 x tr 
\left( PgPg \right)
+   \int d^2 x tr \left( \partial_\tau g\;
P \right)\nonumber \\
&+& { 1\over 2} \int d^2 x tr \left( g^{-1}\partial_\sigma g\; 
g^{-1}\partial_\sigma g \right)
\end{eqnarray}

\noindent and redefine the fields $g$ and $P$ in a way that mimics the
abelian case

\begin{eqnarray}
\label{nonabemand}
g &=& A \; B\nonumber \\
P &=&  \varepsilon\left(B^{-1}\partial_\sigma A^{-1} -\partial_\sigma B^{-1}
A^{-1}\right)
\end{eqnarray}

\noindent  The action for the PCM now reads

\begin{eqnarray}
\label{pcm2}
{\cal S}_{PSM}\left(g,P\right)&=& {\cal S}_{\varepsilon}\left(A\right)+
{\cal S}_{-\varepsilon}\left(B\right)\nonumber\\
& &\mbox{}+ \varepsilon \int d^2x tr\left[A^{-1}
\left(\partial_\tau A \partial_\sigma B - \partial_\sigma A
\partial_\tau B\right)B^{-1}\right]
\end{eqnarray}

\noindent where

\begin{equation}
{\cal S}_{\varepsilon}\left(A\right)=\int d^2x 
tr\left(\varepsilon \partial_\sigma A^{-1}\partial_\tau A 
-\partial_\sigma A^{-1}\partial_\sigma A\right)
\end{equation}

\noindent We see that due to the non-abelian nature of the fields,
the cross-term cannot be eliminated, so that complete separation cannot
be achieved.  This fact should be expected.  In the canonical approach,
we have for their field equations the pair

\begin{eqnarray}
\label{hje}
\partial_\tau I &=& \partial_\sigma J\nonumber\\
\partial_\tau J &=& \partial _\sigma I - \left[I,J\right]
\end{eqnarray}

\noindent where $I= \partial_\tau g\;g^{-1}$ and
$J=\partial_\sigma g\;g^{-1}$. 
The second equation is a sort of Bianchi identity, which is the integrability
condition for the existence of $g$.  Looking at this pair
of equations one can appreciate that chirality is not well defined in this
model.  However this picture changes drastically with the inclusion of the
Wess-Zumino topological term, i.e., when we consider the WZW model. 
The first equation in (\ref{hje}) changes to

\begin{equation}
\partial_\tau I = \partial_\sigma J + \rho \left[I,J\right]
\end{equation}

\noindent with $\rho \in Z$, producing a more symmetric the set of
equations.  In particular for $\rho=\pm 1$ it is known that the
currents above describe two independents affine Lie algebras.  One
expects then that with the introduction of the topological term, one
would be able to obtain an identical mixing term, such that in the
total action they could cancel each other.  It is a lengthy but
otherwise straightforward algebra to show that the topological term

\begin{equation}
\Gamma_{WZ}(g)={1\over 3}\int d^3 x\epsilon^{ijk}tr\left[g^{-1}\partial_i g
\;g^{-1}\partial_j g \;g^{-1}\partial_k g\right]
\end{equation}

\noindent under the field redefinition (\ref{nonabemand}), splits as

\begin{eqnarray}
\label{topol}
\Gamma_{WZ}(AB) &=&\Gamma_{WZ}(A)+\Gamma_{WZ}(B)+\nonumber\\
&+& \int d^2x tr\left[A^{-1}\left(\partial_\tau A \partial_\sigma B 
- \partial_\sigma A
\partial_\tau B\right)B^{-1}\right]
\end{eqnarray}

\noindent  Next, we can bring results (\ref{pcm2}) and (\ref{topol})
into the Wess-Zumino-Witten (WZW) action\cite{Witten}, which is
described by

\begin{equation}
\label{WZW}
{\cal S}_{WZW}(g) = {1\over{\lambda^2}}{\cal S}_{PCM}(g)+{n\over{4\pi}}
\Gamma_{WZ}(g)
\end{equation}

\noindent We mention the appearance of an extra parameter, both in the
action and in the canonical formalism, playing the role of coupling
constant.  In terms of the chiral variables, the WZW model reads

\begin{eqnarray}
\label{wzw2}
{\cal S}_{WZW}\left(A\;B\right)&=&\left[{1\over\lambda^2}
{\cal S}_{\varepsilon}(A)+{n\over 4\pi}\Gamma_{WZ}(A)\right]
+\left[{1\over\lambda^2}{\cal S}_{-\varepsilon}(B)+{n\over 4\pi}
\Gamma_{WZ}(B)\right]+\nonumber\\
&+&  \left({\varepsilon\over\lambda^2}+{n\over 4\pi}\right)
\int d^2x tr\left[A^{-1}\left(\partial_\tau A\partial_\sigma B 
-\partial_\sigma A\partial_\tau B\right)B^{-1}\right]
\end{eqnarray}

\noindent  We can appreciate that the separability condition is only
achieved at the critical points, as expected, but also that our choice
of $\varepsilon$ is now dependent on which of the critical points we
choose: $4\pi\varepsilon=- \lambda^2 n$.
The result is the non abelian version of the chiral decomposition, and
corresponds to the
sum of two Lagrangians describing  non abelian chiral bosons of
opposite chiralities, each one having the form proposed by
Sonnenschein\cite{S}. We will see that a change of the critical point
automatically switches the chirality of $A$ and $B$ by changing the
sign of $\varepsilon$.  Indeed, in order to obtain separability, we
must have either

\begin{equation}
\label{sepcon1}
(i)\:\:\:\:\:\:\:{{\lambda^2n}\over{4\pi}}=-\varepsilon = 1
\end{equation}

\noindent or

\begin{equation}
\label{sepcon2}
(ii)\:\:\:\:\:\:\:{{\lambda^2n}\over{4\pi}}=-\varepsilon = -1
\end{equation}

\noindent In the first case we find the set of chiral equations as

\begin{eqnarray}
\partial_x\left(A^{-1}\partial_+A\right)&=&0\nonumber\\
\partial_-\left(B^{-1}\partial_xB\right)&=&0
\end{eqnarray}

\noindent whose solution reads

\begin{eqnarray}
A&=&A_-(x^-)h_A(t)\nonumber\\
B&=&h_B(t)B_+(x^+)
\end{eqnarray}

In the second case, the chiral equations are

\begin{eqnarray}
\partial_x\left(A^{-1}\partial_-A\right)&=&0\nonumber\\
\partial_+\left(B^{-1}\partial_xB\right)&=&0
\end{eqnarray}

and the solutions read

\begin{eqnarray}
A&=&A_+(x^+)h_A(t)\nonumber\\
B&=&h_B(t)B_-(x^-)
\end{eqnarray}

\noindent The arbitrary functions of time $h_A(t)$ and $h_B(t)$
represent gauge degrees of freedom in the chiral modes, not present
in the original action.  Note that in both cases the general solution
for $g(x^+,x^-)$ is given as

\begin{equation}
g=A_\varepsilon(x^\varepsilon)h_A(t)h_B(t)B_{-\varepsilon}(x^{-\varepsilon})
\end{equation}

\noindent  As in the abelian case, the constraint

\begin{equation}
h_A(t)=h_B^{-1}(t)
\end{equation}

\noindent  becomes necessary in order that $g=AB$ satisfies the
equation of motion for the WZW model.  This constraint is the only
memory left for the chiral bosons stating that they belong to the
same non-chiral field.

%\section{Coupling to gravity}
Now we are in position to consider the coupling of non-abelian chiral
scalars to gravity.  The coupling of a scalar non abelian field to gravity
can be described by the action 

\begin{equation}
{\cal S}_{WZW}(g,\eta) = {1\over{\lambda^2}}{\cal S}_{PCM}(g,\eta)
+{n\over{4\pi}}
\Gamma_{WZ}(g)
\end{equation}

\noindent where

\begin{equation}
\label{GWZW}
{\cal S}_{PCM}(g,\eta) = { 1\over 2} \int d^2 x \sqrt{-\eta}\,tr 
\left( \eta^{\mu\nu}\partial_\mu g \partial_\nu g^{-1} \right) 
\end{equation}

\noindent The topological term $\Gamma_{WZ}(g)$ is not affected
by the metric since the volume element times the antisymmetric tensor
makes a covariant scalar.  We can write the action for the Principal Chiral
Model in a first order form as we did before.  After some algebra, we find

\begin{equation}
{\cal L}_{PCM}(g,\eta)=tr\left[P\partial_\tau g 
+{1\over{2\sqrt{-\eta}\eta^{00}}}
\left(PgPg - \partial_\sigma g\partial_\sigma g^{-1}\right)
+\sqrt{-\eta}\eta^{01}P\partial_\sigma g\right]
\end{equation}

\noindent Introducing the chiral variables defined in (\ref{nonabemand})
the action for the
interacting PCM splits in

\begin{eqnarray}
{\cal S}_{PSM}\left(g,P,\eta\right)&=& {\cal S}_{\varepsilon}
\left(A,{\cal G}_\varepsilon\right)+
{\cal S}_{-\varepsilon}\left(B,{\cal G}_{-\varepsilon}\right)\nonumber\\
& &\mbox{}+ \varepsilon \int d^2x Tr\left[A^{-1}
\left(\partial_\tau A \partial_\sigma B -\partial_\sigma A
\partial_\tau B\right)B^{-1}\right]
\end{eqnarray}

\noindent where

\begin{equation}
{\cal S}_{\varepsilon}\left(A , {\cal G}_\varepsilon\right)=\int d^2x 
tr\left(\varepsilon \partial_\sigma A^{-1}\partial_\tau A 
-{\cal G}_\varepsilon 
\partial_\sigma A^{-1}\partial_\sigma A^{-1}\right)
\end{equation}

\noindent and the couplings ${\cal G}_\varepsilon$ are defined in
(\ref{couplings}).  The WZW model now assumes a form analog to
(\ref{wzw2})

\begin{eqnarray}
\label{wzw3}
{\cal S}_{WZW}\left(A\;B\right)&=&\left[{1\over\lambda^2}
{\cal S}_{\varepsilon}(A , {\cal G}_\varepsilon)
+{n\over 4\pi}\Gamma_{WZ}(A)\right]
+\left[{1\over\lambda^2}{\cal S}_{-\varepsilon}(B , {\cal G}_{-\varepsilon})
+{n\over 4\pi}
\Gamma_{WZ}(B)\right]+\nonumber\\
&+&  \left({\varepsilon\over\lambda^2}+{n\over 4\pi}\right)
\int d^2x tr\left[A^{-1}\left(\partial_\tau A\partial_\sigma B 
-\partial_\sigma A\partial_\tau B\right)B^{-1}\right]
\end{eqnarray}

\noindent In particular, we note that the separability conditions,
Eqs.(\ref{sepcon1}) and (\ref{sepcon2}), are
not affected by the presence of the metric.  Thus, at the separability points,
 we get two chiral
non-abelian scalars coupled to gravity in the same way as in the abelian case.

\vspace{0.3cm}

\noindent {\bf 5}  Concluding, we have proposed a way of splitting
the action for the non-abelian boson into two chiral sectors,
including the interacting case.  This chiral decomposition is related
to the holomorphic factorization of the WZW model discussed in
\cite{KZ,w2}. If we go to Euclidean space our light cone coordinates
$x^+$ and $x^-$ will correspond to the conformal coordinates $z$ and
$\overline z$ and the chiral solutions $B_+(x^+) $ and
$A_-(x^-)$ will correspond to holomorphic and antiholomorphic
functions respectively.

 An interesting point for further investigation is the effect of
quantum corrections in this decomposition.  It is known, for the WZW
model, that quantum corrections contribute just to a renormalization
of the action, as can be seen in \cite{dgn} and in the references
quoted there.  In this case, considering the massless nature of the
WZW model at critical point discussed by Witten, it is natural to
expect that quantum corrections associated to our redefinition of
variables will not spoil the chiral decomposition.\vspace{0.3cm}\\

\noindent Acknowledgments. We would like to thank E. C. Marino for
kindly reading the manuscript. One of us (NRFB) would like to thank
M. Asorey and P. van Nieuwenhuizen for important discussions.  CW
thanks the hospitality of the Department of Physics and Astronomy of
University of Rochester, where part of this work was done.  The
authors are partially supported by CNPq, FINEP and FUJB , Brasil.

\end{document}